\documentclass[aps,prl,twocolumn,floats,epsfig,showpacs]{revtex4-1}

\pdfoutput=1

\usepackage{amsmath}
\usepackage{amsfonts}
\usepackage{exscale}
\usepackage{float}
\usepackage{tikz}

\usepackage{bbm}
\usepackage{amssymb}
\usepackage{ifpdf}
\usepackage{color}
\usepackage{graphicx,epsfig}

\newcommand{\N}{{\mathbb{N}}}
\newcommand{\Z}{{\mathbb{Z}}}
\newcommand{\R}{{\mathbb{R}}}

\newcommand{\1}{{\mathbbm{1}}}
\newcommand{\p}{\partial}
\newcommand{\ontopof}[2]{\genfrac{}{}{0pt}{}{#1}{#2}}

\begin{document}

\title{A New Concept for the Momentum of a Quantum Mechanical Particle in a Box}

\author{M.\ H.\ Al-Hashimi and U.-J.\ Wiese$^\dagger$}
\affiliation{$^\dagger$ Albert Einstein Center, Institute for Theoretical 
Physics, University of Bern, 3012 Bern, Switzerland}

\begin{abstract}
For a particle in a box, the operator $- i \p_x$ is not Hermitean. We provide
an alternative construction of a momentum operator $p = p_R + i p_I$, which has 
a Hermitean component $p_R$ that can be extended to a self-adjoint operator, as 
well as an anti-Hermitean component $i p_I$. This leads to a description of 
momentum measurements performed on a particle that is strictly limited to the 
interior of a box.
\end{abstract}

\maketitle

In quantum mechanics physical observables are described by self-adjoint 
operators. The subtle differences between Hermiticity and self-adjointness,
which arise because the Hilbert space is infinite-dimensional, were first 
understood by von Neumann \cite{Neu32}. In particular, in addition to 
Hermiticity, self-adjointness of an operator $A$ requires that its domain 
$D(A)$ coincides with the domain $D(A^\dagger)$ of its adjoint $A^\dagger$ 
\cite{Ree75,Gie00}. The domain of a differential operator is characterized by 
square-integrability conditions on derivatives of the wave functions. For a 
particle in a finite volume, boundary conditions (characterized by self-adjoint 
extension parameters) further restrict the domain \cite{Bal70,Car90,AlH12}. 
Only self-adjointness (and not Hermiticity alone) guarantees a real-valued 
spectrum of eigenvalues and an orthonormal set of corresponding eigenfunctions. 
This is crucial for the correct description of measurements that return the 
eigenvalues of the corresponding operator. It is well-known that, for a 
particle in a 1-d box $[-\tfrac{L}{2},\tfrac{L}{2}]$, the operator $- i \p_x$ 
is not Hermitean, let alone self-adjoint \cite{Bon01}. In this supposedly 
simple quantum mechanical problem \cite{Aro97,Hu99,Gor01}, momentum 
measurements are hence a non-trivial issue \cite{Gar04}. The standard approach
\cite{Coh77} is to treat the problem in the Hilbert space $L^2(\R)$ of 
square-integrable functions over the entire real axis (for which $- i \p_x$ is 
self-adjoint). One then regularizes the problem in the infrared by turning the 
infinite square-well potential into a finite one, $V(x) = V_0$, 
$|x| > \frac{L}{2}$, and finally takes $V_0 \rightarrow \infty$ \cite{Gar04}. 
Here we pursue an alternative approach that is strictly limited to the interior
of the box. We regularize the problem in the ultraviolet by introducing a 
lattice with spacing $a$, and then send $a \rightarrow 0$. This naturally 
leads to the Hilbert space $L^2([-\tfrac{L}{2},\tfrac{L}{2}]^2)$ as the key to 
the solution of this long-standing problem.

For a particle moving along the positive real axis, the operator $A = - i \p_x$ 
(in units were $\hbar = 1$) is not self-adjoint. By partial integration one 
obtains
\begin{equation}
\langle A^\dagger \chi|\Psi\rangle = \langle\chi|A \Psi\rangle = 
\langle A \chi|\Psi\rangle - i \chi(0)^* \Psi(0) \ .
\end{equation}
Hermiticity, hence requires $\chi(0)^* \Psi(0) = 0$. This can be achieved by
restricting the domain $D(A)$ to those wave functions that obey $\Psi(0) = 0$
and whose derivative is square-integrable. Then $\chi(0)$ can still take 
arbitrary values. As a result, the domain of the adjoint operator $A^\dagger$ 
(which acts on $\chi$) remains unrestricted, such that 
$D(A^\dagger) \supset D(A)$. Since $D(A^\dagger) \neq D(A)$, although $A$ is
Hermitean, it is not self-adjoint. For a particle in a finite interval, the
operator $A = - i \p_x$ is self-adjoint only if one imposes periodic boundary
conditions on the probability density. This is not of interest here, because we
are interested in an interval with physically distinct endpoints.

We now consider a particle in a box with the Hamiltonian
$H = - \tfrac{1}{2 m} \p_x^2 + V(x)$, $x \in [-\tfrac{L}{2},\tfrac{L}{2}]$,
assuming that the potential $V(x)$ is non-singular. Performing two partial
integrations one now obtains
\begin{eqnarray}
&&\langle H^\dagger \chi|\Psi\rangle = \langle\chi|H \Psi\rangle = \nonumber \\ 
&&\langle H \chi|\Psi\rangle + \frac{1}{2 m} 
\left[\p_x \chi(x)^* \Psi(x) - \chi(x)^* \p_x \Psi(x)\right]_{-L/2}^{L/2} \ .
\label{HHermiticity}
\end{eqnarray}
The domain $D(H)$ contains those wave functions whose second derivative is
square-integrable and that obey the Robin boundary conditions
\begin{equation}
\gamma_+ \Psi(\tfrac{L}{2}) + \p_x \Psi(\tfrac{L}{2}) = 0, \
\gamma_- \Psi(-\tfrac{L}{2}) - \p_x \Psi(-\tfrac{L}{2}) = 0.
\label{Robinbc}
\end{equation}
Inserting these relations into the square bracket in eq.(\ref{HHermiticity}), 
the Hermiticity condition takes the form 
\begin{eqnarray}
&&\left[\p_x \chi(\tfrac{L}{2})^* + \gamma_+ \chi(\tfrac{L}{2})^*\right] 
\Psi(\tfrac{L}{2}) - \nonumber \\
&&\left[\p_x \chi(- \tfrac{L}{2})^* - \gamma_- \chi(-\tfrac{L}{2})^*\right] 
\Psi(-\tfrac{L}{2}) = 0 \ .
\end{eqnarray}
Since $\Psi(\pm \tfrac{L}{2})$ can take arbitrary values, this implies
\begin{equation}
\gamma_+^* \chi(\tfrac{L}{2}) + \p_x \chi(\tfrac{L}{2}) = 0 \ , \quad
\gamma_-^* \chi(-\tfrac{L}{2}) - \p_x \chi(-\tfrac{L}{2}) = 0 \ ,
\end{equation}
which characterizes the domain $D(H^\dagger)$ of $H^\dagger$ (that acts on 
$\chi$). The two domains coincide,
$D(H^\dagger) = D(H)$, only if $\gamma_\pm^* = \gamma_\pm \in \R$. This defines
a 2-parameter family of self-adjoint extensions of $H$. The boundary conditions 
of eq.(\ref{Robinbc}) guarantee that the probability current
\begin{equation}
j(x) = \frac{1}{2 m i} [\Psi(x)^* \p_x \Psi(x) - \p_x \Psi(x)^* \Psi(x)] \ ,
\end{equation}
does not leak outside the box, i.e.\ $j(\pm\tfrac{L}{2}) = 0$. Not 
surprisingly, self-adjointness ensures probability conservation. We conclude 
that the wave function need not vanish at the boundary. Dirichlet boundary 
conditions, $\Psi(\pm\tfrac{L}{2}) = 0$, correspond to 
$\gamma_\pm \rightarrow \infty$, while Neumann boundary conditions, 
$\p_x \Psi(\pm\tfrac{L}{2}) = 0$, result from $\gamma_\pm = 0$. 

In the absence of a potential ($V(x) = 0$) it is easy to derive the energy
spectrum, $H \psi_l(x) = E_l \psi_l(x)$, $l \in \N$. The states with 
$E_l > 0$ are given by $\psi_l(x) = A \exp(i k x) + B \exp(- i k x)$, 
with the energy quantization condition
\begin{equation}
\exp(2 i k L) = 
\frac{(\gamma_+ - i k)(\gamma_- - i k)}{(\gamma_+ + i k)(\gamma_- + i k)} \ ,
\quad E_l = \frac{k^2}{2 m} \ .
\label{Encontinuum}
\end{equation}
For negative values of $\gamma_\pm$ there are, in addition, 
negative energy states localized on the boundaries \cite{AlH12}. With general
Robin boundary conditions, parity symmetry requires $\gamma_- = \gamma_+$ in
addition to $V(-x) = V(x)$.

The subtleties associated with Hermiticity versus self-adjointness arise 
because the Hilbert space is infinite-dimensional. In order to gain a better 
understanding of the problem, we first discretize it onto a finite lattice of 
$N$ points with lattice spacing $a$, such that the Hilbert space becomes 
$N$-dimensional \cite{San76,Jag81,Jag82,Tor03}. As illustrated in 
Fig.\ref{Fig1}, we divide the interval $[-\tfrac{L}{2},\tfrac{L}{2}]$ into 
$N = L/a$ segments of size $a$, and introduce a lattice point in the middle of 
each segment, such that $x = n a$, 
$n \in \left\{-\tfrac{N-1}{2},-\tfrac{N-3}{2},\dots,
\tfrac{N-3}{2},\tfrac{N-1}{2}\right\}$. We choose $N$ to be odd, such that 
$n \in \Z$. The kinetic energy is then represented by a discretized second 
derivative, such that the $N \times N$ matrix $H$ takes the form
\begin{eqnarray}
H&=&- \frac{1}{2 m a^2} \left(\begin{array}{ccccccc}
-1 & 1 & 0 & \dots & 0 & 0 & 0 \\
 1 &-2 & 1 & \dots & 0 & 0 & 0 \\
 0 & 1 &-2 & \dots & 0 & 0 & 0 \\
 . & . & . & \dots & . & . & . \\
 . & . & . & \dots & . & . & . \\
 0 & 0 & 0 & \dots &-2 & 1 & 0 \\
 0 & 0 & 0 & \dots & 1 &-2 & 1 \\
 0 & 0 & 0 & \dots & 0 & 1 &-1 \end{array}\right) \nonumber \\
&+&\mbox{diag}\left(V_{-(L-a)/2} + \tfrac{\gamma_-}{2 m a},V_{-(L-3a)/2},
V_{-(L-5a)/2},\right. \nonumber \\ 
&\dots&\left. ,V_{(L-5a)/2},V_{(L-3a)/2},V_{(L-a)/2} + \tfrac{\gamma_+}{2 m a}\right)
\ .
\end{eqnarray}
On the lattice, the self-adjoint extension parameters $\gamma_\pm$ are directly 
incorporated into the Hamiltonian as additional terms on the diagonal, which 
are equivalent to $\delta$-function potentials at the boundary in the continuum 
limit $a \rightarrow 0$. In the absence of a potential ($V_x = 0$) it is
straightforward to solve the Schr\"odinger equation 
$H \psi_{l,x} = E_l \psi_{l,x}$ by the same ansatz as in the continuum, 
$\psi_{l,x} = A \exp(i k x) + B \exp(- i k x)$. The quantization condition for
$E_l = \tfrac{1}{2 m} (\tfrac{2}{a} \sin\tfrac{k a}{2})^2$ takes the form
\begin{eqnarray}
&&\exp(2 i k (L-a)) =
\frac{\gamma_+ + \tfrac{1}{a}[1 - \exp(i k a)] - 2 m E_l a}
{\gamma_+ + \tfrac{1}{a}[1 - \exp(- i k a)] - 2 m E_l a} \nonumber \\
&&\times \frac{\gamma_- + \tfrac{1}{a}[1 - \exp(i k a)] - 2 m E_l a}
{\gamma_- + \tfrac{1}{a}[1 - \exp(- i k a)] - 2 m E_l a} \ .
\label{Enlattice}
\end{eqnarray}
Indeed, this reduces to eq.(\ref{Encontinuum}) in the continuum limit.

On the lattice the momentum operator should be represented by a discretized
first derivative. It is necessary to distinguish forward and backward 
derivatives \cite{Mar16,Mar17,Mar18}
\begin{equation}
p_F = - \frac{i}{a} \left(\begin{array}{ccccccc}
-1 & 1 & 0 & \dots & 0 & 0 & 0 \\
 0 &-1 & 1 & \dots & 0 & 0 & 0 \\
 0 & 0 &-1 & \dots & 0 & 0 & 0 \\
 . & . & . & \dots & . & . & . \\
 . & . & . & \dots & . & . & . \\
 0 & 0 & 0 & \dots &-1 & 1 & 0 \\
 0 & 0 & 0 & \dots & 0 &-1 & 1 \\
 0 & 0 & 0 & \dots & 0 & 0 &\lambda_+ \end{array}\right) \ , \nonumber 
\end{equation}
\begin{figure}[tbp]
\vspace{1cm}
\begin{tikzpicture}
\tikz\draw[line width=0.4mm,color=black] (0,0) -- (8.1,0)
[line width=0.4mm,color=black] (0,-0.3) -- (0,0.3)
[line width=0.4mm,color=black] (8.1,-0.3) -- (8.1,0.3)
[fill=black] (0.45,0) circle (0.5ex)
[fill=black] (1.35,0) circle (0.5ex)
[fill=black] (2.25,0) circle (0.5ex)
[fill=black] (3.15,0) circle (0.5ex)
[fill=black] (4.05,0) circle (0.5ex)
[fill=black] (4.95,0) circle (0.5ex)
[fill=black] (5.85,0) circle (0.5ex)
[fill=black] (6.75,0) circle (0.5ex)
[fill=black] (7.65,0) circle (0.5ex)
[line width=0.4mm,color=black] (0,-0.5) -- (0.45,-0.5)
node[below] {\hspace{-0.45cm}$a/2$}
[line width=0.4mm,color=black] (0,-0.6) -- (0,-0.4)
[line width=0.4mm,color=black] (0.45,-0.6) -- (0.45,-0.4)
[line width=0.4mm,color=black] (4.05,-0.5) -- (4.95,-0.5) 
node[below] {\hspace{-0.9cm}$a$}
[line width=0.4mm,color=black] (4.05,-0.6) -- (4.05,-0.4)
[line width=0.4mm,color=black] (4.95,-0.6) -- (4.95,-0.4);
\end{tikzpicture}
\vspace{-1.5cm}
\caption{\it Lattice with $N = 9$ points in the interval 
$[-\tfrac{L}{2},\tfrac{L}{2}]$.}
\vspace{-1.1cm}
\label{Fig1}
\end{figure}
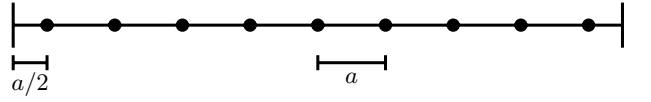
\begin{equation}
p_B = - \frac{i}{a} \left(\begin{array}{ccccccc}
-\lambda_-& 0 & 0 & \dots & 0 & 0 & 0 \\
-1 & 1 & 0 & \dots & 0 & 0 & 0 \\
 0 &-1 & 1 & \dots & 0 & 0 & 0 \\
 . & . & . & \dots & . & . & . \\
 . & . & . & \dots & . & . & . \\
 0 & 0 & 0 & \dots & 1 & 0 & 0 \\
 0 & 0 & 0 & \dots &-1 & 1 & 0 \\
 0 & 0 & 0 & \dots & 0 &-1 & 1 \end{array}\right) \ .
\end{equation}
Just as $\gamma_\pm$, the parameters $\lambda_\pm \in i \R$ will turn into two 
self-adjoint extension parameters in the continuum limit. Since neither $p_F$ 
nor $p_B$ is Hermitean, we construct the Hermitean and anti-Hermitean
combinations
\begin{eqnarray}
p_R&=&\frac{1}{4}(p_F + p_F^\dagger + p_B + p_B^\dagger) \nonumber \\
&=&- \frac{i}{2a} \left(\begin{array}{ccccccc}
-\lambda_- & 1 & 0 & \dots & 0 & 0 & 0 \\
-1 & 0 & 1 & \dots & 0 & 0 & 0 \\
 0 &-1 & 0 & \dots & 0 & 0 & 0 \\
 . & . & . & \dots & . & . & . \\
 . & . & . & \dots & . & . & . \\
 0 & 0 & 0 & \dots & 0 & 1 & 0 \\
 0 & 0 & 0 & \dots &-1 & 0 & 1 \\
 0 & 0 & 0 & \dots & 0 &-1 &\lambda_+ \end{array}\right) \ , \nonumber \\
i p_I&=&\frac{1}{4}(p_F - p_F^\dagger + p_B - p_B^\dagger) \nonumber \\
&=&\frac{i}{2a} \mbox{diag}(1,0,0,\dots,0,0,-1), \ p = p_R + i p_I .
\label{latticemomentumoperators}
\end{eqnarray}
The momentum operator $p$ is not Hermitean, but has a Hermitean component 
$p_R$ and an anti-Hermitean component $p_I$, which is diagonal in the position 
basis. In the continuum limit, it reduces to $\delta$-functions at the 
boundary \cite{Rob63,Rob66}. The Hermitean component $p_R$ results from a 
symmetrized forward-backward next-to-nearest neighbor derivative, which extends 
over two lattice spacings.

A characteristic feature of momentum is that it changes sign under parity. 
Hence, the operator $p$ should anti-commute with the unitary transformation 
$U_P$, with $U_P \Psi_x = \Psi_{-x}$, that represents parity in Hilbert space. 
For $\lambda_+ = \lambda_-$ one indeed obtains $U_P p_F U_P^\dagger = - p_B$, 
$U_P p_B U_P^\dagger = - p_F$, $U_P p_R U_P^\dagger = - p_R$, 
$U_P p_I U_P^\dagger = - p_I$.
It is straightforward to solve the eigenvalue problem 
$p_R \phi_{k,x} = \tfrac{1}{a} \sin(k a) \phi_{k,x} = \hat k \phi_{k,x}$. The 
corresponding eigenstates $\phi_{k,x}$ (with $x = n a$) take the form
\begin{eqnarray}
&&\phi_{k,x} = A \exp(i k x) + B \exp(- i k x) \ , \ 
\mbox{for $n$ even} \ , 
\nonumber \\
&&\phi_{k,x} = A \exp(i k x) - B \exp(- i k x) \ , \ 
\mbox{for $n$ odd} \ ,
\label{latticeansatz}
\end{eqnarray}
and the momentum quantization condition is given by
\begin{equation}
\exp(2 i k L) = \frac{(1 + \lambda_+ \exp(i k a))(1 - \lambda_- \exp(i k a))}
{(\exp(i k a) - \lambda_+)(\exp(i k a) + \lambda_-)}.
\label{latticemomentumquantization}
\end{equation}
For $\lambda_+ = \lambda_- = \pm i$ this implies
\begin{equation}
\label{latticemomenta}
k = \frac{\pi n}{L}, \
n \in \{-\tfrac{N-1}{2},-\tfrac{N-3}{2},\dots,\tfrac{N-3}{2},\tfrac{N-1}{2}\} 
\ .
\end{equation}
It is important to point out that the eigenvalues 
$\hat k = \tfrac{1}{a} \sin(k a)$ are in one-to-one correspondence with the 
values of $k$ from eq.(\ref{latticemomenta}), and are hence not degenerate.

At this stage, we have reached a completely satisfactory description of the 
momentum operator in the $N$-dimensional Hilbert space of the $N$-point 
lattice. However, it is not entirely trivial to obtain an equivalent continuum
description that agrees with the lattice results in the limit 
$a \rightarrow 0$. In this context, it is important to note that the 
eigenfunctions $\phi_{k,x}$ of eq.(\ref{latticeansatz}) depend explicitly on 
whether the lattice point $x = n a$ has an even or odd value of $n$. In fact,
this is the crucial insight that leads to the appropriate mathematical 
description also directly in the continuum. In order to describe the operator
$p_R$ in the continuum, it is necessary to maintain the concept of even and odd 
degrees of freedom. Obviously, the points $x \in \R$ cannot be divided into an 
even and an odd subset. However, it is natural to introduce a two-component 
wave function on which $p_R$ acts as a $2 \times 2$ matrix
\begin{equation}
p_R = - i \left(\begin{array}{cc} 0 & \p_x \\ \p_x & 0 
\end{array}\right) = - i \sigma_1 \p_x, \
\Psi(x) = \left(\begin{array}{c} \Psi_e(x) \\ \Psi_o(x) \end{array}\right).
\label{2component}
\end{equation}
By partial integration we obtain
\begin{eqnarray}
&&\langle p_R^\dagger \chi|\Psi\rangle = \langle \chi|p_R \Psi\rangle = 
\nonumber \\
&&\langle p_R \chi|\Psi\rangle -
i [\chi_e(x)^* \Psi_o(x) + \chi_o(x)^* \Psi_e(x)]_{-L/2}^{L/2} \ .
\label{pRHermiticity}
\end{eqnarray}
We now impose the boundary conditions
\begin{equation}
\Psi_o(\tfrac{L}{2}) = \lambda_+ \Psi_e(\tfrac{L}{2}) \ , \quad
\Psi_o(-\tfrac{L}{2}) = \lambda_- \Psi_e(-\tfrac{L}{2}) \ ,
\label{momentumbc}
\end{equation}
which constrain the domain $D(p_R)$. Inserting these relations in the square 
bracket in eq.(\ref{pRHermiticity}), the Hermiticity condition takes the form
\begin{eqnarray}
&&[\chi_e(\tfrac{L}{2})^* \lambda_+ +  \chi_o(\tfrac{L}{2})^*] 
\Psi_e(\tfrac{L}{2}) - \nonumber \\
&&[\chi_e(-\tfrac{L}{2})^* \lambda_- +  \chi_o(-\tfrac{L}{2})^*] 
\Psi_e(-\tfrac{L}{2}) = 0 \ .
\end{eqnarray}
Since $\Psi_e(\pm \tfrac{L}{2})$ can take arbitrary values, this implies
\begin{equation}
\chi_o(\tfrac{L}{2}) = - \lambda_+^* \chi_e(\tfrac{L}{2}) \ , \quad
\chi_o(-\tfrac{L}{2}) = - \lambda_-^* \chi_e(-\tfrac{L}{2}) \ .
\end{equation} 
Self-adjointness of $p_R$ requires $D(p_R^\dagger) = D(p_R)$, which implies
$\lambda_\pm = - \lambda_\pm^*$ such that $\lambda_\pm \in i \R$.
Hence, there is a 2-parameter family of self-adjoint extensions, characterized 
by the purely imaginary parameters $\lambda_+$ and $\lambda_-$. 

Let us now consider the eigenvalue problem of the self-adjoint operator $p_R$.
In analogy to eq.(\ref{latticeansatz}) on the lattice, we make the ansatz
\begin{equation}
\phi_k(x) = \left(\begin{array}{c} A \exp(i k x) + B \exp(- i k x) \\
A \exp(i k x) - B \exp(- i k x) \end{array}\right) \ .
\label{continuumansatz}
\end{equation}
Imposing the boundary conditions implies
\begin{equation}
\left(\begin{array}{cc} 
(1 - \lambda_+) \exp(i k L) & - (1 + \lambda_+) \\
(1 - \lambda_-) \exp(- i k L) & - (1 + \lambda_-) 
\end{array}\right) \left(\begin{array}{c} A \\ B \end{array}\right) = 0 \ .
\end{equation}
A non-trivial solution arises only when the determinant of the matrix vanishes
\begin{equation}
\exp(2 i k L) = 
\frac{(1 + \lambda_+)(1 - \lambda_-)}{(1 - \lambda_+)(1 + \lambda_-)} \ .
\label{continuummomentumquantization}
\end{equation}
This agrees with the lattice momentum quantization condition of 
eq.(\ref{latticemomentumquantization}) in the continuum limit 
$a \rightarrow 0$. Hence, the 2-component formulation of eq.(\ref{2component}) 
indeed provides the correct continuum description of the momentum operator 
$p_R$ that was constructed on the lattice in 
eq.(\ref{latticemomentumoperators}).

Next we consider the Hamiltonian
\begin{equation}
H(\mu) = \left(\begin{array}{cc} - \tfrac{1}{2 m} \p_x^2 + V(x) & 0 \\ 
0 & - \tfrac{1}{2 m} \p_x^2 + V(x) \end{array}\right) + \mu P_- \ .
\label{doubledHamiltonian}
\end{equation}
Here $P_-$ is a projection operator on states $\Psi^-(x)$ with 
$\Psi^-_o(x) = - \Psi^-_e(x)$. On the lattice, such states have energies at the 
cut-off scale $1/a$. For $\mu \rightarrow \infty$ those are removed from the 
energy spectrum of the continuum theory. The operator $P_+$ projects on the 
remaining finite-energy states $\Psi^+(x)$ with $\Psi^+_o(x) = \Psi^+_e(x)$, 
i.e.\
\begin{eqnarray}
&&P_\pm = \frac{1}{2} 
\left(\begin{array}{cc} 1 & \pm 1 \\ \pm 1 & 1 \end{array}\right) \ , \quad
P_\pm^2 = P_\pm \ , \quad P_+ + P_- = \1 \ , \nonumber \\
&&\Psi(x) = \Psi^+(x) + \Psi^-(x) \ , \quad \Psi^\pm(x) = P_\pm \Psi(x) \ .
\end{eqnarray}
Next we introduce the boundary conditions
\begin{equation}
\left(\begin{array}{c} \Psi_o(\pm \tfrac{L}{2}) \\ 
\p_x \Psi_o(\pm \tfrac{L}{2}) \end{array}\right) = e^{i \theta_\pm}
\left(\begin{array}{cc} a_\pm & \pm b_\pm \\ \pm c_\pm & d_\pm 
\end{array}\right) \left(\begin{array}{c} \Psi_e(\pm \tfrac{L}{2}) \\ 
\p_x \Psi_e(\pm \tfrac{L}{2}) \end{array}\right) \ .
\label{generalbc}
\end{equation}
Again demanding that $j(\pm \tfrac{L}{2}) = 0$, it is easy to derive the 
conditions $a_\pm, b_\pm, c_\pm, d_\pm \in \R$ and 
$a_\pm d_\pm - b_\pm c_\pm = - 1$. Together with $\theta_\pm$, these parameters 
define a family of self-adjoint extensions with eight independent parameters
\cite{Jor13}. However, we are not interested in the most general Hamiltonian in 
this class. We just need to find an appropriate continuum formulation for the 
original Hamiltonian in the Hilbert space of 2-component wave functions, which 
is essential for describing the momentum operator $p_R$. Since the boundary 
conditions should support the finite-energy states with 
$\Psi^+_o(x) = \Psi^+_e(x)$, this implies
\begin{equation}
e^{i \theta_\pm} = 1 \ , \quad a_\pm = 1 \ , \quad b_\pm = 0 \ , \quad
d_\pm = - 1 \ .
\label{specialparameters}
\end{equation}
Using the parameters of eq.(\ref{specialparameters}), eq.(\ref{generalbc}) 
reduces to
\begin{equation}
- \frac{c_\pm}{2} \Psi^+(\pm\tfrac{L}{2}) \pm \p_x \Psi^+(\pm\tfrac{L}{2}) = 0 
\ , \ \Psi^-(\pm\tfrac{L}{2}) = 0 \ .
\end{equation}
When we identify $\Psi^+(x)$ with the wave functions in the original Hilbert
space and set $\gamma_\pm = - c_\pm/2$, this is equivalent to the Robin 
boundary conditions of eq.(\ref{Robinbc}). Interestingly, the wave functions
$\Psi^-(x)$ obey Dirichlet boundary conditions. It is important to note that
the Hamiltonian $H(\mu)$ and the momentum operator $p_R$ are not defined in the
same domains, i.e.\ $D(H(\mu)) \neq D(p_R)$. Since $\lambda_\pm \in i \R$, 
eq.(\ref{momentumbc}) is indeed incompatible with the boundary conditions 
associated with eq.(\ref{specialparameters}). This implies
that an energy eigenstate is a superposition of infinitely many momentum
eigenstates. Furthermore, although at the formal level of differential 
expressions the kinetic energy $T = - \tfrac{1}{2m} \p_x^2$ and the momentum 
$p_R = - i \sigma_1 \p_x$ seem to commute, in fact $[T,p_R] \neq 0$ due to 
domain incompatibilities. An exception are Dirichlet boundary conditions,
$- c_\pm \rightarrow \infty$, because for them 
$\Psi_e(\pm\tfrac{L}{2}) = \Psi_o(\pm\tfrac{L}{2}) = 0$. 

We have now succeeded to represent the physics, which was first constructed on 
the lattice, directly in the continuum. The crucial steps consisted in the 
2-component realization of the physical Hilbert space, which is necessary to 
construct $p_R$ as a self-adjoint operator, and in the elimination of the 
additional states $\Psi^-(x)$ from the spectrum of the Hamiltonian $H(\mu)$ in 
the limit $\mu \rightarrow \infty$.

We can now address the question of momentum measurements for a particle that 
is confined inside a box. The momentum operator $p = p_R + i p_I$ is not 
Hermitean because it also contains the anti-Hermitean 
component 
\begin{equation}
p_I = \frac{1}{2} \lim_{\epsilon \rightarrow 0^+} \left(\begin{array}{cc}
\delta(x + \tfrac{L}{2} - \epsilon) - \delta(x - \tfrac{L}{2} + \epsilon) & 0
\\ 0 & 0 \end{array}\right) \ ,
\end{equation}
which is diagonal in the position basis. Here we have assumed that 
$\tfrac{N-1}{2}$ is even. Since $[p_R,p_I] \neq 0$, the components $p_R$ and 
$p_I$ are not simultaneously measurable. Instead of the usual Heisenberg 
algebra $[x,- i \p_x] = i$ underlying canonical quantization, here we have
operators $A_-(x) = \sin\tfrac{\pi x}{L} + i \sigma_1 \cos\tfrac{\pi x}{L}$ and 
$A_+(x) = A_-(x)^\dagger$ with
\begin{equation}
[p_R,A_\pm] = \pm \frac{\pi}{L} A_\pm\ , \ [A_+,A_-] = 0 \ , \ A_+ A_-= \1 \ .  
\end{equation}
Since $A_+(\pm \tfrac{L}{2}) = A_-(\pm \tfrac{L}{2}) = \pm \1$ the momentum 
shift operators $A_\pm(x)$ do not lead out of the domain $D(p_R)$.

Finally, let us consider the standard textbook case with Dirichlet boundary 
conditions, $\Psi(\pm \tfrac{L}{2}) = 0$, which correspond to the 
parity-symmetric case $\gamma_+ = \gamma_- \rightarrow \infty$, and
\begin{eqnarray}
&&H(\mu) \psi_l(x) = E_l \psi_l(x) \ , \quad E_l = \frac{\pi^2 l^2}{2 m L^2} \ , 
\quad l \in \N_{>0} \nonumber \\ 
&&\psi_l(x) = \frac{1}{\sqrt{L}} 
\left(\begin{array}{c} \cos(\pi l x/L) \\ \cos(\pi l x/L)
\end{array}\right) \ , \quad \mbox{$l$ odd} \ , \nonumber \\ 
&&\psi_l(x) = \frac{1}{\sqrt{L}} 
\left(\begin{array}{c} \sin(\pi l x/L) \\ \sin(\pi l x/L)
\end{array}\right) \ , \quad \mbox{$l$ even} \ . 
\end{eqnarray}
Let us assume that there is no explicit parity breaking due to the momentum.
For $\lambda_+ = \lambda_- = \lambda$ the corresponding momentum eigenvalues 
and eigenfunctions are
\begin{eqnarray}
&&p_R \phi_k(x) = k \phi_k(x), \ k = \frac{\pi n}{L}, \ n \in \Z, \ 
\sigma = \frac{1 - \lambda}{1 + \lambda} \in U(1), \nonumber \\ 
&&\phi_k(x) = \frac{1}{2 \sqrt{L}}
\left(\begin{array}{c} \exp(i k x) + \sigma \exp(- i k x) 
\\ \exp(i k x) - \sigma \exp(- i k x) \end{array}\right), \mbox{$n$ even} \ ,
\nonumber \\ 
&&\phi_k(x) = \frac{1}{2 \sqrt{L}} 
\left(\begin{array}{c} \exp(i k x) - \sigma \exp(- i k x) \\ 
\exp(i k x) + \sigma \exp(- i k x) \end{array}\right), \mbox{$n$ odd.}
\end{eqnarray}
In this case, one obtains 
$\langle \psi_l|p_R|\psi_l\rangle = \langle \psi_l|p_I|\psi_l\rangle = 0$. 
Independent of $\lambda$, when projected onto the finite-energy sector, the 
momentum eigenstates are just 
$\phi^+_{k,e}(x) = \phi^+_{k,o}(x) = \frac{1}{2 \sqrt{L}} \exp(i k x)$. When 
one measures the momentum $p_R$ in an energy eigenstate $\psi_l$, one obtains 
$k = \pm \pi l/L$ each with probability $\tfrac{1}{4}$. The probability to 
measure $k = \pi n/L$ for $n \neq \pm l$ is 
$|\langle \phi_k|\psi_l\rangle|^2 = \tfrac{4}{\pi^2} l^2/(l^2 - n^2)^2$ if 
$(-1)^n = - (-1)^l$ and zero otherwise. The resulting momentum uncertainty is
$\Delta k = \pi l/L$. It is a peculiarity of Dirichlet boundary conditions
that $E_l = \frac{1}{2 m} (\Delta k)^2$. 

For Neumann boundary conditions (i.e.\ $\gamma_\pm = 0$) the ground state is 
$\psi_0(x) = \frac{1}{\sqrt{2 L}} (\ontopof{1}{1})$. Then the probability to 
measure $k = 0$ is $\tfrac{1}{2}$, and to measure $k = \pi n/L$ with $n \neq 0$ 
is $|\langle \phi_k|\psi_0\rangle|^2 = \tfrac{2}{\pi^2 n^2}$ for odd $n$ and 
zero otherwise. For Neumann boundary conditions the momentum uncertainty 
diverges for any energy eigenstate.

After a momentum measurement, the particle is in an eigenstate $\phi_k$, 
with both a $\phi^+_k$ and $\phi^-_k$ component. The energy of the $\phi^-_k$
component diverges for $\mu \rightarrow \infty$. This implies that a momentum 
measurement transfers an infinite amount of energy to the particle. The same 
happens for a particle that moves along the entire real axis in a potential 
that diverges at infinity, $V(\pm \infty) \rightarrow \infty$. Then the 
standard momentum operator $- i \p_x$ is self-adjoint. A measurement that 
yields the momentum $k$ puts the particle in the momentum eigenstate 
$\langle x|k\rangle = \exp(i k x)$, which has $\langle k|V|k\rangle = \infty$. 
When applied to a particle in a box, such a momentum measurement catapults the
particle outside of the box. The Fourier transform 
$\widetilde\psi_l(k) = \int_{-\infty}^\infty dx \ \psi_l(x) \exp(-i k x)$ of the 
wave function for the box with Dirichlet boundary conditions yields the 
probability density $\tfrac{1}{2 \pi}|\widetilde\psi_l(k)|^2$ for measuring the 
standard unquantized momentum $k \in \R$. Interestingly, one again obtains
$\Delta k = \pi l /L$, while with Neumann boundary conditions the uncertainty
again diverges.

Complementary to the operator $- i \p_x$ that acts in $L^2(\R)$, the new 
concept for the momentum of a particle, strictly confined inside a box, 
operates in $L^2([-\tfrac{L}{2},\tfrac{L}{2}]^2)$. Both momentum measurements 
transfer an infinite energy to the particle, the former in the infrared and the 
latter in the ultraviolet. Practical momentum measurements are not described 
exactly by a mathematical idealization. If after a practical momentum 
measurement the particle stays inside the box, its momentum is quantized and 
the new concept should be applied. 

It is easy to generalize $p_R$ to higher dimensions
\begin{equation}
\vec p_R = - i \sigma_1 \vec \nabla \ , \quad 
\Psi_o(\vec x) = \lambda(\vec x) \Psi_e(\vec x) \ , \quad 
\vec x \in \p \Omega \ . 
\end{equation}
Here $\lambda(\vec x) \in i \R$ where $\vec x$ belongs to the boundary of a 
finite region $\Omega \in \R^d$ with impenetrable walls. It will be interesting 
to further explore the physical implications of the new momentum concept, e.g., 
in the context of the Heisenberg uncertainty relation in a finite volume 
\cite{AlH12}.

We like to thank M.\ Blau and C.\ Tretter for very interesting discussions.

\end{document}